\documentclass[pre,showpacs,amsmath,amssymb,showkey,preprint,titlepage]{revtex4}
\usepackage{epsfig}
\usepackage{graphicx}% Include figure files
\usepackage{dcolumn}% Align table columns on decimal point
\usepackage{bm}% bold math
\usepackage[T1]{fontenc}
\usepackage[latin1]{inputenc}
\usepackage{array}
\usepackage{longtable}
\usepackage{setspace}
\usepackage{multind}

\begin{document}

\title{SIMULATING GROVER'S QUANTUM SEARCH IN A CLASSICAL COMPUTER}
\author{D.K. Ningtyas and A.B. Mutiara}
\email{amutiara@staff.gunadarma.ac.id}
\homepage{http://stafsite.gunadarma.ac.id/amutiara}
\affiliation{Graduate Program in Information System, Gunadarma
University, \\
Jl. Margonda Raya 100, Depok 16424, Indonesian.}

%\date{Submitted 5 May 2007}

\begin{abstract}
The rapid progress of computer science has been accompanied by a
corresponding evolution of computation, from classical computation
to quantum computation. As quantum computing is on its way to
becoming an established discipline of computing science, much
effort is being put into the development of new quantum
algorithms. One of quantum algorithms is Grover algorithm, which
is used for searching an element in an unstructured list of N
elements with quadratic speed-up over classical algorithms. In
this work, Quantum Computer Language (QCL) is used to make a
Grover's quantum search simulation in a classical computer
\end{abstract}
%\pacs{61.20.Ja; 61.20c.Lc; 64.70.Pf}
%\doi{99900}

\maketitle 

\section{INTRODUCTION}

\subsection{Background of the Research}

The rapid progress of computer science has been accompanied by a
corresponding evolution of computation, from classical computation
to quantum computation. In classical computation, computer memory
made up of bits, where each bit represents either a one or a zero.
In quantum computation, there are some quantum mechanical
phenomena, such as superposition and entanglement, to perform
operations on data.

Instead of using bits, quantum computation uses qubits (quantum
bits). A single qubit can represent a one, a zero, or both at the
same time, which is called superposition. Because of this ability,
quantum computation can perform many tasks simultaneously, faster
than classical computing. There is also another phenomenon in
quantum computation which is called entanglement. If two qubits
get an outside force, then those qubits can be entangled
condition. It means that, even the distance of both qubits is far,
treating one of them will affect the other qubit too. For example,
there are two entangled qubits, and one of them has spin up (we
know it after done a measurement). Then without have to measure
it, we can directly know that the other qubit has spin down.
Because of this ability, communication in quantum computation can
reach a very high speed because information can be transferred
instantly, very fast like it overmatch the speed of light.

As quantum computing is on its way to becoming an established
discipline of computing science, much effort is being put into the
development of new quantum algorithms. One of quantum algorithms
is Grover algorithm, which is used for searching an element in an
unstructured list of N elements with quadratic speed-up over
classical algorithms. Today, there are some quantum programming
languages which can be used to simulate quantum mechanical and
quantum algorithm without having a real quantum computer. In this
work, Quantum Computer Language (QCL) will be used to make a
Grover's quantum search simulation in a classical computer.

\subsection{Position of the Research}

This research is related to an invention of a quantum search
algorithm by Lov K. Grover \cite{Grover_1996}. His invention presents an algorithm,
which is known as Grover algorithm, that is significantly faster
than any classical algorithm can be. This quantum search algorithm
can search for an element in an unsorted database containing N
elements only in \emph{O}($\sqrt{\emph{N}}$) steps, while in the
models of classical computation, searching an unsorted database
cannot be done in less than linear time (so merely searching
through every item is optimal), which will be done in
\emph{O}(\emph{N}) steps. Also, this research is related with
Paramita et al work, where their paper presented a pseudo code for
better understanding about Grover algorithm, and Freddy P. Zen et
al work \cite{Zen_2003}, who provide an example simulation of Grover algorithm in
their paper. This research will also try to simulate Grover
algorithm in a classical computer using one of quantum programming
languages, Quantum Computer Language (QCL)\cite{Oemer_1996,Oemer_2003}.

\subsection{Significance of the Research}

In practice, this research can be used as the fastest known method
or solution for searching an element in an unsorted database
containing N elements. By using the method in this research, the
searching process can speed-up quadratically over classical
algorithms.

\subsection{Problem}

Considered points in this work are:
\begin{enumerate}
\item Is it possible to simulate Grover algorithm in a classical
computer? \item How many qubits and iterations the program needed
to search an element? \item How minimum and maximum the size of
elements in the database that the program can hold?
\end{enumerate}

\subsection{Objective}

The objective of this work is to to make a simulation of Grover
algorithm using Quantum Computer Language (QCL), to know how many
qubits and iterations needed for the searching process, and to
know how minimum and maximum the size of elements in the database
that can be hold by the program.

\subsection{Scope}

This thesis concern on simulating Grover algorithm in a classical
computer using Quantum Computer Language (QCL). The program can
search a desired element in an unsorted database of N elements.

\subsection{Working Methodology}

This work begins with designing pseudo code and flowchart for
Grover algorithm. Then, the design will be implemented by using
Quantum Computer Language. After that, there will be several test
to know how many qubits and iterations needed for the searching
process, also to know how minimum and maximum the size of elements
in the database that can be hold by the program.
 %Bab 1

\section{LITERATURE REVIEW}

Computer science has grown faster, made an evolution in
computation. Research has already begun on what comes after our
current computing revolution. This research has discovered the
possibility for an entirely new type of computer, one that
operates according to the laws of quantum physics - a quantum
computer.

\subsubsection{Way to Quantum Computation}
Quantum computers were first proposed in the 1970s and 1980s by
theorists such as Richard Feynman, Paul Benioff, and David
Deutsch. At that times, many scientists doubted that they could
ever be made practical. Richard Feynman was the first to suggest,
in a talk in 1981, that quantum-mechanical systems might be more
powerful than classical computers. In this lecture
\cite{Feynmann_1982}, reproduced in the International Journal of
Theoretical Physics in 1982, Feynman asked what kind of computer
could simulate physics and then argued that only a quantum
computer could simulate quantum physics efficiently. He focused on
quantum physics rather than classical physics. He said that nature
isn't classical, and if we want to make a simulation of nature,
we'd better make it quantum mechanical, because it does not look
so easy. Around the same time, in a paper titled "Quantum
mechanical models of Turing machines that dissipate no energy"
\cite{Benioff_1982} and related articles, Paul Benioff
demonstrated that quantum-mechanical systems could model Turing
machines. In other words, he proved that quantum computation is at
least as powerful as classical computation. But is quantum
computation more powerful than classical computation? David
Deutsch explored this question and more in his 1985 paper "Quantum
theory, the Church-Turing principle and the universal quantum
computer" \cite{Deutsch_1985}. First, he introduced quantum
counterparts to both the Turing machine and the universal Turing
machine. He then demonstrated that the universal quantum computer
can do things that the universal Turing machine cannot, including
generate genuinely random numbers, perform some parallel
calculations in a single register, and perfectly simulate physical
systems with finite dimensional state spaces. In 1989, in "Quantum
computational networks" \cite{Deutsch_1989}, Deutsch described a
second model for quantum computation: quantum circuits. He
demonstrated that quantum gates can be combined to achieve quantum
computation in the same way that Boolean gates can be combined to
achieve classical computation. He then showed that quantum
circuits can compute anything that the universal quantum computer
can compute, and vice versa.

\subsection{Quantum Computers Development} Quantum computers could
one day replace silicon chips, just like the transistor once
replaced the vacuum tube. But for now, the technology required to
develop such a quantum computer is beyond our reach. Most research
in quantum computing is still very theoretical.

The most advanced quantum computers have not gone beyond
manipulating more than 16 qubits, meaning that they are a far cry
from practical application. However, the potential remains that
quantum computers one day could perform, quickly and easily,
calculations that are incredibly time-consuming on conventional
computers. Several key advancements have been made in quantum
computing in the last few years. Let's look at a few of the
quantum computers that have been developed.

\begin{itemize}
\item In 1998, Los Alamos and MIT researchers managed to spread a
single qubit across three nuclear spins in each molecule of a
liquid solution of alanine (an amino acid used to analyze quantum
state decay) or trichloroethylene (a chlorinated hydrocarbon used
for quantum error correction) molecules. Spreading out the qubit
made it harder to corrupt, allowing researchers to use
entanglement to study interactions between states as an indirect
method for analyzing the quantum information. \item In March 2000,
scientists at Los Alamos National Laboratory announced the
development of a 7-qubit quantum computer within a single drop of
liquid. The quantum computer uses nuclear magnetic resonance (NMR)
to manipulate particles in the atomic nuclei of molecules of
trans-crotonic acid, a simple fluid consisting of molecules made
up of six hydrogen and four carbon atoms. The NMR is used to apply
electromagnetic pulses, which force the particles to line up.
These particles in positions parallel or counter to the magnetic
field allow the quantum computer to mimic the information-encoding
of bits in digital computers. Researchers at IBM-Almaden Research
Center developed what they claimed was the most advanced quantum
computer to date in August. The 5-qubit quantum computer was
designed to allow the nuclei of five fluorine atoms to interact
with each other as qubits, be programmed by radio frequency pulses
and be detected by NMR instruments similar to those used in
hospitals (see How Magnetic Resonance Imaging Works for details).
Led by Dr. Isaac Chuang, the IBM team was able to solve in one
step a mathematical problem that would take conventional computers
repeated cycles. The problem, called order-finding, involves
finding the period of a particular function, a typical aspect of
many mathematical problems involved in cryptography. \item In
2005, the Institute of Quantum Optics and Quantum Information at
the University of Innsbruck announced that scientists had created
the first qubyte, or series of 8 qubits, using ion traps. \item In
2006, Scientists in Waterloo and Massachusetts devised methods for
quantum control on a 12-qubit system. Quantum control becomes more
complex as systems employ more qubits. \item In 2007, Canadian
startup company D-Wave demonstrated a 16-qubit quantum computer.
The computer solved a sudoku puzzle and other pattern matching
problems. The company claims it will produce practical systems by
2008. Skeptics believe practical quantum computers are still
decades away, that the system D-Wave has created isn't scaleable,
and that many of the claims on D-Wave's Web site are simply
impossible (or at least impossible to know for certain given our
understanding of quantum mechanics).
\end{itemize}

If functional quantum computers can be built, they will be
valuable in factoring large numbers, and therefore extremely
useful for decoding and encoding secret information. If one were
to be built today, no information on the Internet would be safe.
Our current methods of encryption are simple compared to the
complicated methods possible in quantum computers. Quantum
computers could also be used to search large databases in a
fraction of the time that it would take a conventional computer.
Other applications could include using quantum computers to study
quantum mechanics, or even to design other quantum computers.

But quantum computing is still in its early stages of development,
and many computer scientists believe the technology needed to
create a practical quantum computer is years away. Quantum
computers must have at least several dozen qubits to be able to
solve real-world problems, and thus serve as a viable computing
method.

\subsection{Superposition}
Superposition is the fundamental law of quantum mechanics. It
defines the collection of all possible states that an object can
have. Superposition means a system can be in two or more of its
states simultaneously. For example a single particle can be
travelling along two different paths at once.

The principle of superposition states that if the world can be in
any configuration, any possible arrangement of particles or
fields, and if the world could also be in another configuration,
then the world can also be in a state which is a superposition of
the two, where the amount of each configuration that is in the
superposition is specified by a complex number.

For example, if a particle can be in position A and position B, it
can also be in a state where it is an amount "3i/5" in position A
and an amount "4/5" in position B. To write this, physicists
usually say:

\begin{center}$|\psi\rangle = {3\over 5} i |A\rangle + {4\over 5}
|B\rangle$\end{center}

In the description, only the relative size of the different
components matter, and their angle to each other on the complex
plane. This is usually stated by declaring that two states which
are a multiple of one another are the same as far as the
description of the situation is concerned.

\begin{center}$|\psi \rangle \approx \alpha |\psi \rangle$\end{center}

The fundamental dynamical law of quantum mechanics is that the
evolution is linear, meaning that if the state A turns into A' and
B turns into B' after 10 seconds, then after 10 seconds the
superposition $\psi$ turns into a mixture of A' and B' with the
same coefficients as A and B.

\subsubsection{Example}
A particle can have any position, so that there are different
states which have any value of the position x. These are written:

\begin{center}$|x\rangle$\end{center}

The principle of superposition guarantees that there are states
which are arbitrary superpositions of all the positions with
complex coefficients:

\begin{center}$\sum_x \psi(x) |x\rangle$\end{center}

This sum is defined only if the index x is discrete. If the index
is over $\mathbb{R}$, then the sum is not defined and is replaced
by an integral instead. The quantity $\psi(x)$ is called the
wavefunction of the particle.

If a particle can have some discrete orientations of the spin, say
the spin can be aligned with the z axis $|+\rangle$ or against it
$|-\rangle$, then the particle can have any state of the form:

\begin{center}$C_1 |+\rangle + C_2 |-\rangle$\end{center}

If the particle has both position and spin, the state is a
superposition of all possibilities for both:

\begin{center}$\sum_x \psi_+(x)|x,+\rangle + \psi_-(x)|x,-\rangle$\end{center}

The configuration space of a quantum mechanical system cannot be
worked out without some physical knowledge. The input is usually
the allowed different classical configurations, but without the
duplication of including both position and momentum.

A pair of particles can be in any combination of pairs of
positions. A state where one particle is at position x and the
other is at position y is written $|x,y\rangle$. The most general
state is a superposition of the possibilities:

\begin{center}$\sum_{xy} A(x,y) |x,y\rangle$\end{center}

The description of the two particles is much larger than the
description of one particle; it is a function in twice the number
of dimensions. This is also true in probability, when the
statistics of two random things are correlated. If two particles
are uncorrelated, the probability distribution for their joint
position P(x,y) is a product of the probability of finding one at
one position and the other at the other position:

\begin{center}$P(x,y) = P_x (x) P_y(y)$\end{center}

In quantum mechanics, two particles can be in special states where
the amplitudes of their position are uncorrelated. For quantum
amplitudes, the word entanglement replaces the word correlation,
but the analogy is exact. A disentangled wavefunction has the
form:

\begin{center}$A(x,y) = \psi_x(x)\psi_y(y)$\end{center}

while an entangled wavefunction does not have this form. Like
correlation in probability, there are many more entangled states
than disentangled ones. For instance, when two particles which
start out with an equal amplitude to be anywhere in a box have a
strong attraction and a way to dissipate energy, they can easily
come together to make a bound state. The bound state still has an
equal probability to be anywhere, so that each particle is equally
likely to be everywhere, but the two particles will become
entangled so that wherever one particle is, the other is too.

\subsection{Entanglement}
Quantum entanglement, also called the quantum non-local
connection, is a property of a quantum mechanical state of a
system of two or more objects in which the quantum states of the
constituting objects are linked together so that one object can no
longer be adequately described without full mention of its
counterpart - even if the individual objects are spatially
separated in a spacelike manner. The property of entanglement was
understood in the early days of quantum theory, although not by
that name. Quantum entanglement is at the heart of the EPR paradox
developed in 1935. This interconnection leads to non-classical
correlations between observable physical properties of remote
systems, often referred to as nonlocal correlations.

Quantum mechanics holds that observable, for example, spin are
indeterminate until such time as some physical intervention is
made to measure the observable of the object in question. In the
singlet state of two spins it is equally likely that any given
particle will be observed to be spin-up as that it will be
spin-down. Measuring any number of particles will result in an
unpredictable series of measures that will tend more and more
closely to half up and half down. However, if this experiment is
done with entangled particles the results are quite different. For
example, when two members of an entangled pair are measured, their
spin measurement results will be correlated. Two (out of
infinitely many) possibilities are that the spins will be found to
always have opposite spins (in the spin anti-correlated case), or
that they will always have the same spin (in the spin correlated
case). Measuring one member of the pair therefore tells you what
spin the other member would have if it were also measured. The
distance between the two particles is irrelevant.

Theories involving 'hidden variables' have been proposed in order
to explain this result; these hidden variables account for the
spin of each particle, and are determined when the entangled pair
is created. It may appear then that the hidden variables must be
in communication no matter how far apart the particles are, that
the hidden variable describing one particle must be able to change
instantly when the other is measured. If the hidden variables stop
interacting when they are far apart, the statistics of multiple
measurements must obey an inequality (called Bell's inequality),
which is, however, violated - both by quantum mechanical theory
and in experiments.

When pairs of particles are generated by the decay of other
particles, naturally or through induced collision, these pairs may
be termed "entangled", in that such pairs often necessarily have
linked and opposite qualities, i.e. of spin or charge. The
assumption that measurement in effect "creates" the state of the
measured quality goes back to the arguments of, among others:
Schrödinger, and Einstein, Podolsky, and Rosen concerning
Heisenberg's uncertainty principle and its relation to observation
(see also the Copenhagen interpretation). The analysis of
entangled particles by means of Bell's theorem, can lead to an
impression of non-locality (that is, that there exists a
connection between the members of such a pair that defies both
classical and relativistic concepts of space and time). This is
reasonable if it is assumed that each particle departs the scene
of the pair's creation in an ambiguous state (as per a possible
interpretation of Heisenberg). In such a case, for a given
measurement either outcome remains a possibility; only measurement
itself would precipitate a distinct value. On the other hand, if
each particle departs the scene of its "entangled creation" with
properties that would unambiguously determine the value of the
quality to be subsequently measured, then a postulated
instantaneous transmission of information across space and time
would not be required to account for the result. The Bohm
interpretation postulates that a guide wave exists connecting what
are perceived as individual particles such that the supposed
hidden variables are actually the particles themselves existing as
functions of that wave.

Observation of wavefunction collapse can lead to the impression
that measurements performed on one system instantaneously
influence other systems entangled with the measured system, even
when far apart. Yet another interpretation of this phenomenon is
that quantum entanglement does not necessarily enable the
transmission of classical information faster than the speed of
light because a classical information channel is required to
complete the process.

\subsection{Hadamard Transform}
The Hadamard transform (also known as the Walsh-Hadamard
transform, Hadamard-Rademacher-Walsh transform, Walsh transform,
or Walsh-Fourier transform) is an example of a generalized class
of Fourier transforms. It is named for the French mathematician
Jacques Solomon Hadamard, the German-American mathematician Hans
Adolph Rademacher, and the American mathematician Joseph Leonard
Walsh. It performs an orthogonal, symmetric, involutional, linear
operation on 2m real numbers (or complex numbers, although the
Hadamard matrices themselves are purely real).

The Hadamard transform can be regarded as being built out of
size-2 discrete Fourier transforms (DFTs), and is in fact
equivalent to a multidimensional DFT of size
$2\times2\times\cdots\times2\times2$. It decomposes an arbitrary
input vector into a superposition of Walsh functions.

The Hadamard transform Hm is a $2m × 2m$ matrix, the Hadamard
matrix (scaled by a normalization factor), that transforms 2m real
numbers xn into 2m real numbers Xk. The Hadamard transform can be
defined in two ways: recursively, or by using the binary (base-2)
representation of the indices n and k.

Recursively, we define the $1 × 1$ Hadamard transform H0 by the
identity $H0 = 1$, and then define Hm for $m > 0$ by:

$H_m = \frac{1}{\sqrt2} \begin{pmatrix} H_{m-1} & H_{m-1} \\
H_{m-1} & -H_{m-1} \end{pmatrix}$,

where the $1/\sqrt2$ is a normalization that is sometimes omitted.
Thus, other than this normalization factor, the Hadamard matrices
are made up entirely of 1 and -1.

Equivalently, we can define the Hadamard matrix by its (k, n)-th
entry by writing

$k=k_{m-1} 2^{m-1} + k_{m-2} 2^{m-2} + \cdots + k_1 2 + k_0$,

and

$n=n_{m-1} 2^{m-1} + n_{m-2} 2^{m-2} + \cdots + n_1 2 + n_0$,

where the kj and nj are the binary digits (0 or 1) of n and k,
respectively. In this case, we have:

$\left( H_m \right)_{k,n} = \frac{1}{2^{m/2}} (-1)^{\sum_j k_j
n_j}$.

This is exactly the multidimensional
$2\times2\times\cdots\times2\times2$ DFT, normalized to be
unitary, if the inputs and outputs are regarded as
multidimensional arrays indexed by the $n_j$ and $k_j$,
respectively. Some examples of the Hadamard matrices follow.

$H_0 = +1$

$H_1 = \frac{1}{\sqrt2} \begin{pmatrix}\begin{array}{rr} 1 & 1 \\
1 & -1 \end{array}\end{pmatrix}$

(This H1 is precisely the size-2 DFT. It can also be regarded as
the Fourier transform on the two-element additive group of Z/(2).)

$H_2 = \frac{1}{2} \begin{pmatrix}\begin{array}{rrrr} 1 & 1 & 1 &
1\\ 1 & -1 & 1 & -1 \\ 1 & 1 & -1 & -1 \\ 1 & -1 & -1 &
1\end{array}\end{pmatrix}$

$H_3 = \frac{1}{2^{3/2}} \begin{pmatrix}\begin{array}{rrrrrrrr} 1
& 1 & 1 & 1 & 1 & 1 & 1 & 1\\ 1 & -1 & 1 & -1 & 1 & -1 & 1 & -1 \\
1 & 1 & -1 & -1 & 1 & 1 & -1 & -1 \\ 1 & -1 & -1 & 1 & 1 & -1 & -1
& 1 \\ 1 & 1 & 1 & 1 & -1 & -1 & -1 & -1\\ 1 & -1 & 1 & -1 & -1 &
1 & -1 & 1 \\ 1 & 1 & -1 & -1 & -1 & -1 & 1 & 1 \\ 1 & -1 & -1 & 1
& -1 & 1 & 1 & -1\end{array}\end{pmatrix}$.

$(H_n)_{i,j} = \frac{1}{2^{n/2}} (-1)^{i \cdot j}$

where i $\cdot$ j is the bitwise dot product of the binary
representations of the numbers i and j. For example, $H_{32} =
(-1)^{3 \cdot 2} = (-1)^{(1,1) \cdot (1,0)} = (-1)^{1+0} = (-1)^1
= -1$ , agreeing with the above (ignoring the overall constant).
Note that the first row, first column of the matrix is denoted by
$H_{00}$. The rows of the Hadamard matrices are the Walsh
functions.

In quantum information processing the Hadamard transformation,
more often called Hadamard gate, is a one-qubit rotation, mapping
the qubit-basis states $|0 \rangle$  and $|1 \rangle$  to two
superposition states with equal weight of the computational basis
states $|0 \rangle$  and $|1 \rangle$. Usually the phases are
chosen so that we have

\begin{center}$\frac{|0\rangle+|1\rangle}{\sqrt{2}}\langle0|+\frac{|0\rangle-|1\rangle}{\sqrt{2}}\langle1|$\end{center}

in Dirac notation. This corresponds to the transformation matrix

\begin{center}$H_1=\frac{1}{\sqrt{2}}\begin{pmatrix} 1 & 1 \\ 1 & -1
\end{pmatrix}$\end{center}

in the $|0 \rangle$, $|1 \rangle$ basis.

Many quantum algorithms use the Hadamard transform as an initial
step, since it maps n qubits initialized with $|0 \rangle$ to a
superposition of all 2n orthogonal states in the $|0 \rangle$, $|1
\rangle$ basis with equal weight.

Hadamard gate operations:

$H|1\rangle =
\frac{1}{\sqrt{2}}|0\rangle-\frac{1}{\sqrt{2}}|1\rangle$.

$H|0\rangle =
\frac{1}{\sqrt{2}}|0\rangle+\frac{1}{\sqrt{2}}|1\rangle$.

$H( \frac{1}{\sqrt{2}}|0\rangle-\frac{1}{\sqrt{2}}|1\rangle )=
\frac{1}{2}( |0\rangle+|1\rangle) - \frac{1}{2}( |0\rangle -
|1\rangle) = |1\rangle$;

$H( \frac{1}{\sqrt{2}}|0\rangle+\frac{1}{\sqrt{2}}|1\rangle )=
\frac{1}{\sqrt{2}} \frac{1}{\sqrt{2}}(|0\rangle + |1\rangle) +
\frac{1}{\sqrt{2}}(
\frac{1}{\sqrt{2}}|0\rangle-\frac{1}{\sqrt{2}}|1\rangle)=
|0\rangle$.

\subsection{Grover's Quantum Search Algorithm}
One of the most celebrated achievement of quantum computation is
Lov Grover's quantum search algorithm (known as Grover's
algorithm), which was invented in 1996. Grover's algorithm is a
quantum algorithm for searching an unsorted database with N
entries in O(N1/2) time and using O(log N) storage space.

In models of classical computation, searching an unsorted database
cannot be done in less than linear time (so merely searching
through every item is optimal). Grover's algorithm illustrates
that in the quantum model searching can be done faster than this;
in fact its time complexity O(N1/2) is asymptotically the fastest
possible for searching an unsorted database in the quantum model.
It provides a quadratic speedup.

There are already related works about Grover algorithm, such as
done by S. Paramita et al, Matthew Whitehead, Ahmed Younes, and C.
Lavor et al. S. Paramita et al wrote a pseudo code for Grover
algorithm in their paper \cite{Paramita}. They also gave the
example of Grover's implementation using their pseudo code.
Grover's algorithm can be be combined with another search
algorithm. Matthew Whitehead's paper \cite{Whitehead_2005} shows
how Grover's quantum search may be used to improve the
effectiveness of traditional genetic search on a classical
computer. He use repeated applications of Grover's Algorithm to
get a variety of decent chromosomes that will then be used to form
a starting population for classical genetic search. He also
provides the pseudo code for the modified genetic search, which is
a combination between Grover's quantum search and standard genetic
search. Another work related to Grover algorithm is done by Ahmed
Younes. In his paper \cite{Younes_2008}, he described the
performance of Grover's algorithm. Also, C. Lavor et al wrote a
review about Grover algorithm by means of a detailed geometrical
interpretation and a worked out example. Some basic concepts of
Quantum Mechanics and quantum circuits are also reviewed.
 %Bab 2

\section{DESIGN AND IMPLEMENTATION }

Many problems in classical computer science can be reformulated as
searching a list for a unique element which matches some
predefined condition. If no additional knowledge about the
search-condition C is available, the best classical algorithm is a
brute-force search i.e. the elements are sequentially tested
against C and as soon as an element matches the condition, the
algorithm terminates. For a list of N elements, this requires an
average of $N/2$ comparisons. By taking advantage of quantum
parallelism and interference, Grover found a quantum algorithm
\cite{Grover_1996} which can find the matching element in only
O($\sqrt N$) steps.

In this thesis, Grover algorithm and its implementation will be
explained process by process. The algorithm consists of two parts:
(1) Input and initialization (2) Main loop. Each of the parts will
be explained and implemented one by one below.

\subsection{Input and Initialization}

\subsubsection{Input}
This simulation needs to know what number it should search, so
user will be prompted to input a round number (integer). The
implementation of this input process can be seen below.

\begin{verbatim}
input "Masukkan bilangan bulat yang ingin dicari:",bil;
\end{verbatim}

In the code implementation above we can see that bil is a variable
that is used to store the round number.

\subsubsection{Initialization}
Initialization is a process to initiate variables and qubit
registers needed in the simulation.

The most important variables that we have to intiate are the
number of qubits and the number of iterations needed. Assume that
the number of qubits is called jmlqubit, and the number of
iterations is called iterasi.

To calculate the number of qubits needed, we can use this formula:

$jmlqubit = \lfloor(\log_2 bil)+1\rfloor$

To calculate the number of iterations needed, we can use this
formula:

$iterasi = \lceil\pi/8*\sqrt{2^{jmlqubit}}\rceil$

Then, after the value of both jmlqubit and iterasi are known,
another important step to do is to set up the registers for each
qubits. Also, some variables need to be listed to for common
process; looping, storing result, etc. The code implementation for
initialization can be seen below.

\begin{verbatim}
    int jmlqubit = floor(log(bil,2))+1;
    int iterasi = ceil(pi/8*sqrt(2^jmlqubit));
    int hasilmeasurement;
    int i;
    qureg q[jmlqubit];
    qureg f[1];
    print "Jumlah qubit yang digunakan:",jmlqubit;
    print "Jumlah iterasi yang dibutuhkan:",iterasi;
    print "Proses pencarian dimulai...";
\end{verbatim}

\subsection{Main Loop}

Main loop is the main process to begin searching. The steps to do
in the main loop are:

\begin{enumerate} \item Reset all qubits to $|0\rangle$ and apply
the Hadamard transform to each of them. \item Repeat the following
operation as much as the number of iterations needed (see the
initialization part):
\begin{itemize} \item Rotate the marked state by a phase of $\pi$
radians ($I^\pi_f$). A query function needs to be applied. The
query function is needed to flip the variable f if x (the qubits)
is equal to 1111... \item Apply a phase process between pi and f.
\item Undo the query function. \item Apply a diffusion function.
The process are apply Hadamard transform, invert q, then apply a
phase process between pi and q (rotate if q=1111..). After that,
undo the invert process and undo Hadamard transform. \item Do an
oracle function by measure the quantum register that has been
found, then compare the result to the input.
\end{itemize}
This iterations must be repeated again if the measurement result
does not match with the wanted number.
\end{enumerate}

The code implementation of the main loop including the functions
in it can be seen below.

\begin{verbatim}
    {
        reset;
        H(q);
        for i= 1 to iterasi {
            print "Iterasi",i;
            query(q,f,bil);
            CPhase(pi,f);
            !query(q,f,bil);
            diffuse(q);
        }
        oracle(q,hasilmeasurement,bil);
    }       until hasilmeasurement==bil;
    reset;
\end{verbatim}

\subsubsection{Query Procedure}
\begin{verbatim}
procedure query(qureg x,quvoid f,int bil) {
    int i;
    for i=0 to #x-1 {
        if not bit(bil,i)
            {Not(x[i]);}
    }
    CNot(f,x);
    for i=0 to #x-1 {
        if not bit(bil,i)
            {!Not(x[i]);}
    }
}
\end{verbatim}

\subsubsection{Diffuse Procedure}
\begin{verbatim}
procedure diffuse(qureg q) {
    H(q);
    Not(q);
    CPhase(pi,q);
    !Not(q);
    !H(q);
}
\end{verbatim}

\subsubsection{Oracle Procedure}
This procedure is for checking whether the measurement result is
match with the wanted number or not. In general, the oracle
function can be formulated as below.

\begin{center}$f(x)=\{\begin{array}{r} 1$ if $x=x_0\\ 0$ if $x \neq x_0\\\end{array}$\end{center}

x is the indexes in the database, and $x_0$ is the wanted index.
Back to the simulation, before we implement the oracle, we need to
do a measurement to check if the number that been found is already
matched with the wanted number. The code implementation can be
seen below.

\begin{verbatim}
procedure oracle(qureg q,int hasilmeasurement,bil) {
    measure q,hasilmeasurement;
    if hasilmeasurement==bil {
        print "Hasil measurement:",hasilmeasurement;
        print "Telah sama dengan bilangan yang dicari...";
    }
    else {
        print "Hasil measurement:",hasilmeasurement;
        print "Belum sama dengan bilangan yang dicari...";
   }
}
\end{verbatim}
 %Bab 3

\section{RESULT AND DISCUSSION}

The grover's quantum search simulation can be running from Linux's
terminal, by going to the directory where the file is put in then
typing "qcl -i -b32 SimulasiGrover.qcl". This command will start
QCL then run a file named SimulasiGrover.qcl, and providing all
qubits that QCL has (32 qubits). 

To discuss the results of the program, table \ref{tab:output}
containing ten outputs from grover's quantum search simulation
program is provided.

\begin{table}[!ht]
    \centering
    \caption{Outputs from the program} \vspace{.2cm}
    \begin{tabular}{|l|l|l|l|l|}
    \hline
    \textbf{Input} & \textbf{Qubits}& \textbf{Iterations}& \textbf{List of Measured Number} & \textbf{Total Iterations}\\
    \hline
    10 & 4 & 2 & 10 & 2\\    \hline
    30 & 5 & 3 & 30 & 3\\    \hline
    175 & 8 & 7 & 175 & 7\\    \hline
    500 & 9 & 9 & 373 - 500 & 18\\    \hline
    1000 & 10 & 13 & 327 - 1000 & 26\\    \hline
    1676 & 11 & 18 & 1676 & 18\\    \hline
    2000 & 11 & 18 & 1645 - 1497 - 1493 - 703 - 2000 & 90\\    \hline
    2200 & 12 & 26 & 3765 - 2349 - 2200 & 78\\    \hline
    8111 & 13 & 36 & 8111 & 36\\    \hline
    9999 & 14 & 54 & 9999 & 54\\    \hline
    \hline
    \end{tabular}
    \label{tab:output}
\end{table}

In table \ref{tab:output}, column "Input" is for the number that
the user wants to find. Column "Qubits" is the total of qubits
needed to search the number. Column "Iterations" is the total
iterations needed to find one number to be measuring. Column "List
of Measured Numbers" is the list of numbers that are found and get
measured until the number is same to the input. Column "Total
Iterations" is the total of iterations needed to find the correct
number. The value of this column is the multiplication of the
value in column "Iterations" and the amount of numbers in column
"List of Measured Number.

From the table, we can see that the number of qubits and the
number of iterations needed are depend on the value of the number
that user wants to find. If the number is bigger, so will the
qubits and the iterations be. Sometimes, the number that the
program found is not matched with the input. If this condition is
happen, the program will do the iterations again until the number
is matched with the input. But even the program do the iterations
more than one round, the total iterations is never exceed the
value of the input. We can see this from the table in the column
"Total Iterations". But this Grover's quantum search simulation
has a limitation, the maximum qubits that the program can use is
only 32 qubits (QCL limitation). For the possible real
implementation, grover algorithm can be used for searching a
record in database and improving the traditional genetic search.
 %Bab 4

\section{CONCLUSION REMARKS }

\subsection{Conclusion}

Using QCL, a Grover's quantum search simulation has been made. It
is performed without using a quantum computer, but using a classic
computer. To search an element, the program needs to use qubit
instead of bit. The number of qubits needed is depend on the value
of the number that we want to find. The bigger the value of the
number, the bigger qubits needed. It goes the same with the number
of iterations needed. The minimum value for the number is 1, and
the maximum value is depend on the qubits needed. At this far, the
program has been tested to search number till 9999. This Grover's
quantum search is just a simulation to simulate the algorithm, not
a real quantum searching program that can be implemented on the
real database.

\subsection{Future Work}

This Grover's quantum search is just a simulation of quantum
search in a classic computer. That are some possible works for the
future related to grover algorithm. Some of them is implementing
grover algorithm in a real database using quantum computer, but in
this case, the database must be converted in to quantum states
which is probably the most difficult thing to do. Another possible
work is improving the traditional genetic search by combine it
with grover algorithm.
 %Bab 5
%\bibliographystyle{IEEEtranS}
\bibliography{reference}
\newpage %lampiran
%\addcontentsline{toc}{chapter}{APPENDIX} \singlespacing
\begin{center}
\begin{large}\textbf{APPENDIX}\\\end{large}
\end{center}
\vspace{5mm}

\normalsize \noindent \textbf{Listing Program}\\
\begin{verbatim}
procedure query(qureg x,quvoid f,int bil) {
    int i;
    for i=0 to #x-1 {     // x -> NOT (x XOR bil)
        if not bit(bil,i)
        {Not(x[i]);}
    }
    CNot(f,x);            // flip f jika x=1111..
    for i=0 to #x-1 {     // x <- NOT (x XOR bil)
        if not bit(bil,i)
            {!Not(x[i]);}
    }
}

procedure diffusi(qureg q) {
    H(q);                 // Transformasi Hadamard (superposisi)
    Not(q);               // Inversi q
    CPhase(pi,q);         // Rotate jika q=1111..
    !Not(q);              // undo inversi
    !H(q);                // undo Transformasi Hadamard
}

procedure algoritma(int bil) {
    int jmlqubit = floor(log(bil,2))+1;         // banyaknya qubit
    int iterasi = ceil(pi/8*sqrt(2^jmlqubit));  // banyaknya iterasi
    int hasilmeasurement;
    int i;
    qureg q[jmlqubit];
    qureg f[1];
    print "Jumlah qubit yang digunakan:",jmlqubit;
    print "Jumlah iterasi yang dibutuhkan:",iterasi;
    print "Proses pencarian dimulai...";
    {
        reset;      // bersihkan register
        H(q); // persiapan superposisi
        for i= 1 to iterasi {     // looping utama
            print "Iterasi",i;
            query(q,f,bil);     // hitung C(q)
            CPhase(pi,f);         // negasi |n>
            !query(q,f,bil);      // undo C(q)
            diffusi(q);
        }
    //oracle
        measure q,hasilmeasurement; // measurement
    if hasilmeasurement==bil {
            print "Hasil measurement:",hasilmeasurement;
            print "Telah sama dengan bilangan yang dicari...";
        }
        else {
            print "Hasil measurement:",hasilmeasurement;
            print "Belum sama dengan bilangan yang dicari...";
        }
    } until hasilmeasurement==bil;
    reset; // bersihkan register
}

procedure mulai(){
    int bil;
    print;
    print "-----------------------------------------";
    print;
    print "SIMULASI PENCARIAN KUANTUM MENGGUNAKAN ALGORITMA GROVER";
    print;
    input "Masukkan bilangan bulat yang ingin dicari:",bil;
    algoritma(bil);
    print;
    print "-----------------------------------------";
}
\end{verbatim}

\end{document}